\documentclass[10pt,preprint]{iopart}

\usepackage{graphicx}
\usepackage{bm}

\def\dd{\mbox{d}}

\def\s{\sigma}

\begin{document}


\title[Translocation driven by random sequential adsorption]{Exact
  steady-state velocity of ratchets driven by random sequential
  adsorption}

\author{Maria R. D'Orsogna$^{1}$, Tom Chou$^{1,2}$, and Tibor Antal$^{3}$}
\address{$^{1}$Department of Mathematics, UCLA, Los Angeles, CA 90095-1555}
\address{$^{2}$Department of Biomathematics, UCLA, Los Angeles, CA 90095-1766}
\address{$^{3}$Program for Evolutionary Dynamics, Harvard University, Cambridge, MA 02138}


\date{\today}

\begin{abstract}
We solve the problem of discrete translocation of a polymer through a
pore, driven by the irreversible, random sequential adsorption of
particles on one side of the pore. Although the kinetics of the wall
motion and the deposition are coupled, we find the exact steady-state
distribution for the gap between the wall and the nearest deposited
particle.  This result enables us to construct the mean translocation
velocity demonstrating that translocation is faster when the adsorbing
particles are smaller.  Monte-Carlo simulations also show that smaller
particles gives less dispersion in the ratcheted motion. We also
define and compare the relative efficiencies of ratcheting by
deposition of particles with different sizes and we describe an
associated ``zone-refinement'' process.
\end{abstract}
\pacs{05.60.-k,87.16.Ac,05.10.Ln}

\vspace{5mm}

\section{Introduction}
Polymer translocation through membrane nanopores is a 
common process in living cells.
The transport of proteins and nucleic acids in and out of organelles
serve a variety of control, signaling, and error correction functions
\cite{OSTER0,OSTER1,ELSTON1}.  Recent advances in polymer
manipulation at the nanoscale level have also sparked interest in pore
translocation as a new tool in genetic sequencing, structure
determination, and drug delivery \cite{gerland, meller,turner}.  
Active polymer transport through pores requires driving
forces which often provided by ``chaperone'' proteins that
bind to the polymer on one side of the membrane.  The proteins are
larger than the pore, and once bound, create a barrier blocking
backward polymer fluctuations. This ratcheting process 
eventually drives the entire polymer through the pore.
Another known translocation mechanism is by ``power-stroke''
\cite{ELSTON1,ZANDI}, where chaperones that are deposited close to the
membrane are subject to conformational changes. These induce a strain
that is relieved only by direct pulling of the protein through the
pore, similar to the driving mechanisms of motors such as myosin and
kinesin \cite{vale}.  In post-translational protein translocation,
both Brownian ratcheting \cite{RAT0,RAT1} and power stroke
\cite{GLICK} pulling, mediated by Hsp-70 ATPases, have been
proposed. Both models exhibit qualitatively similar behavior and
cannot be distinguished by experimental data \cite{ELSTON1}. However,
translocation by power stroke is molecularly more complex, its
modeling requires additional parameters \cite{ELSTON1,KOLO0} and its
effects arise only at extremely high binding protein densities
\cite{ZANDI}.  Thus, we will only consider Brownian ratcheting as the
dominant translocation driving force, and neglect power stroke
mechanisms.

\begin{figure}[t]
\begin{center}
\includegraphics[height=0.85in]{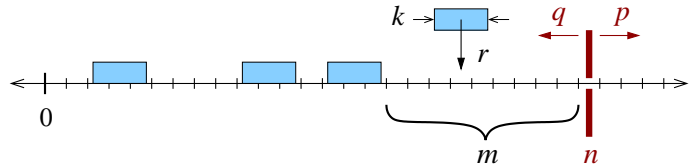}
\end{center}
\caption{Random sequential adsorption of chaperone particles
of $k$ lattice sites, with deposition 
rate $r$. In the polymer reference
frame the fluctuating wall has intrinsic hopping rates
$p,q$ and is ratcheted by the irreversibly bound particles.
In the case of Hsp70 proteins binding to nucleic acids,
$k \sim 6$.}
\label{FIG1}
\end{figure}

The problem of translocating polymers driven by Brownian ratchets has
been considered by many authors. Analytical progress, however, has
been possible only for certain limiting cases where restrictions are
placed on the binding kinetics.  For instance, continuum Fokker-Planck
models have been solved \cite{OSTER0,OSTER1,ELSTON1,ZANDI}
only in the limit of perfect ratchets where particles are forced to
deposit next to the membrane, or in the limit of rapid particle
equilibration (Langmuir kinetics). While the continuum approach is
justifiable in the limit of large chaperone particles that occupy
$k\gg 1$ lattice sites, in many applications this limit may be
unrealistic.  For instance, chaperones of the Hsp70 family, commonly
employed in polymer translocation across the endoplasmic reticulum,
are approximately 2nm in size. If we assume polymers of nucleic acids
with interbase distances of $\sim 0.36$nm, and that the polymer
diffuses one base pair at a time, typical Hsp70 class chaperones would be
described by binding particles with $k \sim 6$. For the translocation
of such structures, the discrete approach is more pertinent.  Random
chaperone particle deposition was recently studied using a discrete
master equation, and including particle detachment and diffusion.
However, results were derived only in the limit of rapid
equilibration, either of the binding particles or of the fluctuating
polymer \cite{METZLER}. In this paper we report an exact steady-state
solution of the discrete translocation process in the irreversible
deposition limit.  In our model, the only constraint is that the
binding particles do not overlap. No other approximations are made.
As in previous work we will only consider stiff polymers that do not
contribute conformational entropy \cite{OSTER1,ZANDI,METZLER}.
The dependence of the mean velocity on the size of the deposited
particles is explicitly computed.  Our result for the mean
translocation velocity is verified using Monte-Carlo simulations.
Simulation results for the dispersion of the translocation is also
discussed.



\section{Solution of Random Sequential Adsorption Ratchet}

Consider particles of integer length $k$ irreversibly binding to an
infinitely long one-dimensional lattice representing the translocating
polymer.  The fluctuating polymer is assumed to jump one unit to the
right or left with rates $q,p$ respectively. In the reference frame of
the polymer, as shown in Fig.\,\ref{FIG1}, the membrane wall hops
forward and backward with rates $p,q$, respectively.  Particle
deposition occurs at rate $r$ only if there are at least $k$ open
sites between the wall and the nearest deposited particle.  The
dynamics of the wall is closely related to that of its nearest gap,
since wall fluctuations are allowed only if there are enough empty
sites for the wall to perform its random motion. Ratcheting occurs
when a gap is large enough for a particle to irreversibly deposit,
preventing the wall from sliding backwards. After deposition,
particles cannot diffuse nor detach.

The master equation for the probability density $P_{m,n}(t)$ for a
wall to be at position $n \in (-\infty,\infty)$ and 
for the gap closest to it to be of
length $m \in [0, \infty)$ can be derived in analogy with problems in
Random Sequential Adsorption (RSA) \cite{evans, privman}.  While in
RSA all gaps are equivalent and one is concerned with the probability
of finding a gap of length $m$ anywhere along the infinite lattice,
here, we are interested in the random deposition of particles in the
\textit{single} gap between the wall and its nearest particle,
as shown in Fig.\,\ref{FIG1}. The time evolution of
$P_{m,n}(t)$ obeys

\begin{equation}
\fl \displaystyle \dot{P}_{m,n} = p P_{m-1,n-1} + q
P_{m+1,n+1}-\left[p+q+r(m -k +1)H_{m-k}\right]P_{m,n} + r\!\!\!\!\sum_{j
  = m + k}^{\infty}\!\!\!\!\!P_{j,n},
\label{ME} 
\end{equation}


\noindent where the Heaviside function $H_{m-k} =1$ for $m\geq k$ and
zero otherwise.  For $m=0$, $\dot{P}_{0,n}(t) = q P_{1,n+1} - p
P_{0,n} + r \sum_{j = k}^{\infty} P_{j,n}$. 
The terms with $p,q$
represent the hopping of a free, non-interacting wall, which by
themselves would lead to a wall drift proportional to $p-q$.  The
other terms in Eq.\,\ref{ME} describe RSA dynamics.  A gap of length
$m$ can be produced by the deposition of a $k$-mer in a gap of
arbitrary length $j \geq m+k$.  Although there are $j-k+1$ ways, each
with rate $r$, of depositing a $k$-mer in such a gap, only one of
these choices will lead to the creation of a gap of length $m$.
Similarly, a gap of length $m$ can be destroyed by the deposition of
any particle of length $k$ within it, a process which occurs in
$(m-k+1)$ ways for an overall destruction rate of $r(m -k +1)$. For
$m<k$, no deposition-mediated gap destruction occurs because a $k$-mer
cannot fit into such a small gap.  Upon summing $P_{m,n}(t)$ over all
values $m \in [0,\infty)$ to define $Q_n(t) \equiv \sum_m P_{n,m}(t)$,
  the terms pertaining to the RSA process cancel exactly and

\begin{equation}
\dot{Q}_{n}= p\left[Q_{n-1}-Q_{n}\right] + q\left[Q'_{n+1} -qQ'_{n}\right],
\label{ME1a}
\end{equation}

\noindent where $Q'_{n} \equiv Q_{n} - P_{0,n}$ is the conditional
probability that the wall is at position $n$ and that the site
preceding it is empty.  Upon multiplying Eq. \ref{ME1a} by $n$ and summing over the 
infinite lattice,

\begin{equation}
{\dd \langle n(t)\rangle \over \dd t} = p-q\!\!\sum_{j=-\infty}^{\infty}\!\!Q'_{j} \equiv p-q\langle\sigma\rangle.
\label{DNDT}
\end{equation}


\noindent where $\langle\sigma\rangle \equiv
\sum_{j=-\infty}^{\infty}Q_{j}'$ is the probability that the site
immediately preceding the wall is empty.  This definition implies that
$\langle\sigma\rangle$, the realization-averaged value of the random
vacancy variable $\sigma$ in the frame of the wall, is also the
probability for a gap of nonzero length to exist between the wall and
the nearest particle.  Provided $Q_{j}'$ reaches its steady-state
distribution in finite time, Eq.  \ref{DNDT} defines the steady-state
mean wall velocity

\begin{equation}
\label{ratch}
v = p-q\langle\sigma\rangle.
\end{equation}



Note that the dependence of the average velocity $v$ on the deposition
rate $r$ resides completely in the term $\langle\sigma\rangle$.  What
remains is to find an explicit steady-state expression for $\langle
\sigma \rangle= \sum_{j}Q'_{j}$.
To this end, we sum Eq.\,\ref{ME}, this time over both wall positions
$n \in (-\infty, \infty)$ and over gap lengths $m' \geq m$ obtaining
the cumulative probability distribution for the first gap to be of
length $m$ or larger $R_m \equiv \sum_{m' = m}^{\infty}\sum_{n =
  -\infty}^{\infty} P_{m,n}(t \to \infty)$.
We may now recognize that $R_1 = \langle\sigma\rangle$, since the
probability for a gap of nonzero length to exist adjacent to the wall
is equivalent to the probability that gap is of {\it any} length
except zero.  Thus, in order to find exact expressions for the
velocity in Eq.\,\ref{ratch} we need to find $R_1$.  Performing the
sums over $n$ and $m'$ in the steady-state limit of Eq.\,\ref{ME}, we
obtain the recursion relation for $R_m$,

\begin{eqnarray}
\fl\:\hspace{1cm} \left[p + q + r(m - k+1)H_{m-k+1}\right]R_{m} = qR_{m+1}+ pR_{m-1}
- r\!\!\!\!\!\!\!\!\!\!\!\!\!\sum_{j=max\{k,m+1\}}^{m+k-1}\!\!\!\!\!\!\!\!\!\!\!\!\!\!R_{j},
\label{tosum}
\end{eqnarray}

\noindent along with the normalization condition $R_{0} =
1$.  Eq.\,\ref{tosum} is solved by introducing the $z$-transform, 

\begin{equation}
G(z)\equiv\! \sum_{m=-\infty}^{\infty} \!z^m R_m,
\end{equation}

\noindent and its inverse 

\begin{equation}
R_{m} =
\oint_C \frac{G(z)} {z^{m+1}}{\dd z \over 2\pi i},
\end{equation}

\noindent given by the Cauchy integral encircling the origin
\cite{jeffreys}.  Although physically $m \in [0,\infty)$, the
$z$-transform is a sum over all integers, extending the definition to
$R_{m<0}$. In order to $z$-transform Eq.\,\ref{tosum} we set
$H_{m-k+1}$ to unity regardless of $m$. In this case, we find a first
order differential equation for $G(z)$ which is solved by

\begin{equation}
G(z) = \frac{G_0z^{k}}{z^{p+q+1}} \exp \left(p z - \frac
{q-H_{k-2}}{z}+ \sum_{j=3}^{k} 
\frac{z^{1-j}}{j-1} \right)
\label{kg3}.
\end{equation}

\noindent For notational simplicity, Eqs. \ref{kg3}-\ref{MATRIX} are
expressed with the rates $p$ and $q$ normalized by the deposition rate
$r$. The $R_{m}$ arising from inverting Eq.\,\ref{kg3} are valid only
for $m\geq k-2$ since the smallest $m$ for which $H_{m-k+1}=1$ is
$m=k-1$ and since, for $m=k-1$, Eq.\,\ref{tosum} contains $R_{k-2}$.
For $k=1,2$ the last term in the exponent of Eq.\,\ref{kg3} vanishes
and the above generating function is valid for all values of $m$. The
Cauchy integral formula thus yields $R_m$ for all values of $m > 0$.
The integration constant $G_{0}$ can be fixed by directly applying the
condition $R_{0}=1$.  For $k=1$ we find



\begin{eqnarray}
\label{k1sol}
R_m &=& \left(\frac {p} q\right)^{m/2}
\frac{J_{p+q+m} (2 \sqrt{p q})}
{J_{p+q} (2 \sqrt{p q})}, \quad\,\, k=1.
\end{eqnarray}
where $J_{\nu}$ is the Bessel function of the first kind of order $\nu$.
For dimers, $G(z)$ is identical to the $k=1$ case except for the
substitution $q \to q-1$ in Eq.\,\ref{kg3}.  For $q > 1$ the form of
$R_m$ is the same as in Eq.\,\ref{k1sol}, with the same substitution
in $q$.  
If we define $Z_{\nu} \equiv J_{\nu}$ for $q>1$ and $Z_{\nu} \equiv I_{\nu}$
for $q<1$, where $I_{\nu}$ is the modified Bessel function of
order $\nu$, for $k=2$ we obtain

\begin{equation}
\label{k2sol}
R_m = \frac{p^{m/2}}{\vert q - 1\vert^{m/2}}
\frac{Z_{p+q+m -1} (2 \sqrt{p\vert q-1\vert})} {Z_{p+q-1} (2
  \sqrt{p\vert q-1\vert})}, \quad k =2.
\end{equation}

\noindent  For $q=1$ in the $k=2$ case, the direct  inverse $z$-transform gives $R_m =
p^{m} \Gamma(p+1)/\Gamma(p+m+1)$, where $\Gamma$ is the
Gamma function.

Next, consider the case of larger particles $k \geq 3$.  The $R_{m\geq
k-2}$ arising from Eq.\,\ref{kg3} must now be coupled with explicit
solutions for $R_{m\leq k-3}$ from Eq.\,\ref{tosum}, where $H_{m-k+1}
=0$.  Let us derive $R_{m\geq k-2}$ from the generating function
$G(z)$.  If we define $a_{j}^{(k)}$ as the $j^{th}$ term in the Laurent
series appearing in Eq.\,\ref{kg3}, $\exp \left( \sum_{j=2}^{k-1}
z^{-j}/j\right) \equiv \sum_{j=0}^{\infty} a_{j}^{(k)} z^{-j}$, we can
write $R_m$ for $m \geq k - 2$ as

\begin{equation}
R_m = G_0 \sum_{j = 0}^{\infty} \,
\frac{a_{j}^{(k)}\, p^{\alpha/2}}{|q-1|^{\alpha/2}}
Z_{\alpha} (2 \sqrt{p \, |q-1|}),
\quad k \geq 3,
\label{nogo}
\end{equation}

\noindent where $\alpha \equiv j+p+q-k+1+m$. These $z$-transformed
solutions are valid only for $m \geq k-2$, for which we let $R_{m}
\equiv G_{0}\tilde{R}_{m}$. In order to apply the condition $R_{0}=1$
and determine $G_{0}$, we must connect Eq.\,\ref{nogo} to the $m \leq k-2$
equations in (\ref{tosum}). These involve $R_{m}$ up to $m=2k-3$.
$G_{0}$ can thus be determined by


\noindent 

\begin{eqnarray}
\label{system}
{\bf M} \cdot
\left( \begin{array}{c}
R_1  \\
\vdots \\
R_{k-3} \\
G_0 \tilde R_{k-2} \\
\vdots \\
G_0 \tilde R_{2k-3}
\end{array} \right)
& = & -
\left( \begin{array}{c}
p   \\
\vdots \\
0 \\
0\\
\vdots \\
0
\end{array} \right).
\end{eqnarray}

\noindent where ${\bf M}$ is the $(2k -3) \times (k-2)$ transition
matrix describing the linear subsystem in Eqs.\,\ref{tosum}:

\begin{equation}
\fl {\bf M} =
\left(\begin{array}{ccccccccc}
-(p+q) & q & 0 & \ldots & 0 & -1 & 0 & 0 & \dots \\
p & -(p+q)& q & \ldots & 0  & -1 & -1 & 0 & \dots \\
0 & p & -(p+q) & \ldots & 0 & -1 & -1 & -1 & \dots\\
\vdots & \vdots & \vdots & \ldots& \vdots & \vdots & \vdots & \vdots & \vdots \\
0 & \dots & 0 & \ldots & q & -1 & -1 & \dots & -1 \\
\end{array}\right) 
\label{MATRIX}
\end{equation}
Solving Eq.\,\ref{system} allows us to determine $G_{0}$ and
the exact gap distribution $R_m$ for all parameters $p,q,r,k$.

\begin{figure*}
\includegraphics[height=5.1cm]{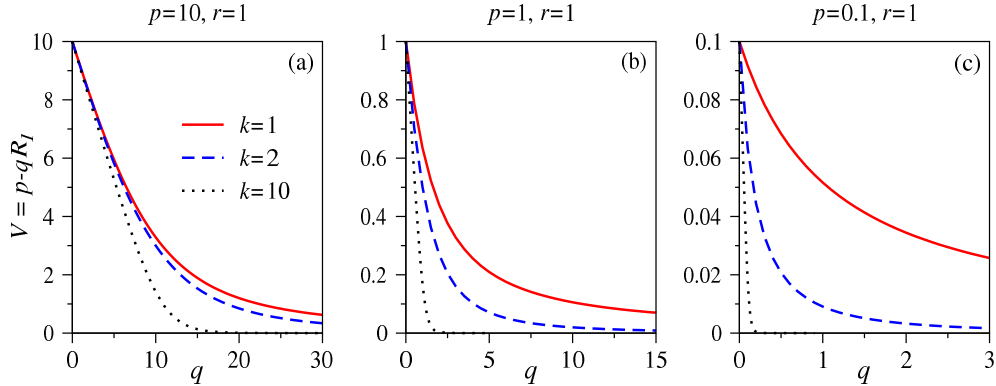}
\caption{(a-c) Exact steady-state velocities as functions of $q$ for $r=1$,
  $p=0.1, 1, 10$ at different particle sizes $k = 1, 2,10$.  Departure
  from a simple biased random walk occurs for large $q/p$. The
  irreversible deposition of particles ensures $v > 0$. For each set
  of values $p,q$, the maximal velocity arises for the smallest
  possible particles, when $k=1$.}
\label{FIG2}
\end{figure*}

\section{Velocity and Dispersion}

In Figs.\,\ref{FIG2}(a-c) we plot the average wall velocity $v = p
-qR_1$ as functions of the backward hopping rate $q$, for various
particle lengths $k$ and forward hopping rates $p$.  Ratcheting from
particle deposition is stronger when $q/p$ is large, the wall motion
is biased towards the deposited particles, and $R_{1}$ is small. For
fixed kinetic parameters, particles of smaller size $k$ yield faster
translocation and smaller variance. Smaller particles are more
effective at translocation due to their enhanced insertion rate into
the fluctuating gap despite taking smaller steps than larger
particles. The relative difference of the velocity and dispersion
among different particles sizes is most pronounced in the strongly
ratcheting regime where $q/p$ is large.

Now consider the dispersion of the ratcheted wall. 
Upon multiplying Eq. \ref{ME1a} by $n^2$ and summing 
over all integers $n$, we find 

\begin{equation}
{\dd \over \dd t}\langle n^{2}\rangle = p+q\langle\sigma\rangle +2p\langle n\rangle 
-2q\!\!\sum_{n=-\infty}^{\infty}\!\!n Q_{n}'.
\end{equation}

\noindent For this calculation, we cannot simply use $\langle n\rangle
= vt$ as implied by Eq. \ref{ratch}.  Although one expects the
realization-averaged $\langle\sigma\rangle$ to exponentially decay to
its steady-state value, an initial transient exists before the
distributions reach steady-state.  Incorporating an ``integration
constant'' arising from this transient, and integrating Eq. \ref{DNDT}
(with $n(0)=0$), we define

\begin{equation}
\langle n(t) \rangle = pt -
q\int_{0}^{t}\langle\s\rangle \dd t \equiv vt + n_{0}.
\end{equation}


\noindent Similarly, we define

\begin{equation}
\displaystyle \langle n'\rangle \equiv {\sum_{n=-\infty}^{\infty} n Q'_{n} \over 
\sum_{n=-\infty}^{\infty} Q'_{n}} \equiv {\sum_{n=-\infty}^{\infty} n Q'_{n} \over 
\langle \sigma\rangle} =  vt + n_{0}'.
\end{equation}

\noindent The two {\it different} integration constants $n_{0}$ and
$n_{0}'$ that embody the initial transients do not affect the
determination of the steady-state velocity $v=\dd \langle
n(t)\rangle/\dd t$, but interestingly, affects the wall dispersion.
Upon computing $\langle n^{2}\rangle - \langle n\rangle^{2} = 2Dt$, we
find

\begin{equation}
2D = p+q\langle\sigma\rangle \left[1-2(n_{0}'-n_{0})\right].
\label{DISPERSION}
\end{equation}

\noindent The dispersion $D$ retains memory of the
transients during which the averaged velocities have not yet reached
$v$. This contribution to the dispersion is embodied in the
$n_{0}'-n_{0}$ term in Eq. \ref{DISPERSION}. The term $n_{0}'-n_{0}$
is always positive because it takes longer for a wall with a gap to
reach terminal velocity than an unrestricted wall, leading to a larger
intercept $n_{0}'$. Therefore $p+q\langle\sigma\rangle$ is an upper
bound for $2D$ that is accurate for $k=1$ in the $r\rightarrow \infty$ limit
where $\langle\sigma\rangle \rightarrow 0$.

\begin{figure*}
\begin{center}
\includegraphics[height=5.4cm]{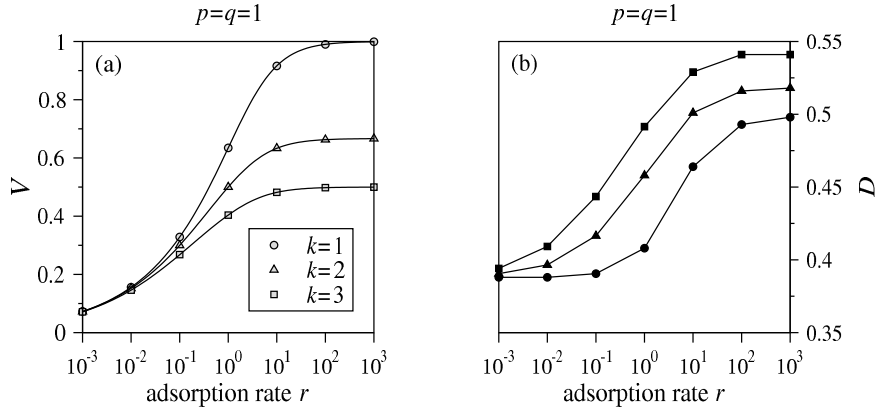}
\end{center}
\caption{Velocities and dispersions as a function of deposition rate
$r$ for fixed $p=q=1$ and $k=1,2,3$. (a) The mean velocity $v$ for
$k=1,2,3$. Both Monte-Carlo results and the exact solution are shown
together.  (b) Monte-Carlo results for the dispersion $D$. For $r\gg
p,q$, the limiting values (from Eq. \ref{VDRSYM}) for $D$ are $1/2,
14/27$, and $13/24$ for $k=1,2$, and $3$, respectively.}
\label{FIG3}
\end{figure*}

In Figs. \ref{FIG3}, we show both the mean velocity $v$ and 
dispersion $D$ of the wall, this time as functions of $r$, with fixed  $p=q=1$.
As $r$ is increased, not only does the mean velocity $v$ increase, but
so does the dispersion $D$.  Although the wall is ``pushed'' harder by
the rapidly deposited particles, its typical displacement also
increases to slightly overcompensate the sharpening effect of imposing
reflecting boundary conditions at each deposited particle. In the
$r\rightarrow 0$ limit, we expect the dispersion to approach that of a
diffusing particle with reflecting boundary conditions on one side:
$D=p(1-2/\pi) \approx 0.3634p$ for $q=p$. The Monte-Carlo value for
$p=q=k=1$ and $r=10^{-3}$ gives $D=0.388$ in good agreement with the
expected result.

The exact result for the mean velocity given by Eq. \ref{ratch} and
$R_{1}$ from Eq. \ref{k1sol} or \ref{k2sol} can be simplified in the
slow deposition rate limit $r\rightarrow 0$ corresponding to low
chaperone concentrations.  For example, for $k=1$ and $q/r = p/r
\rightarrow \infty$, asymptotic expressions for $R_{1}$ can be found
using the recursion relation

\begin{equation}
J_{\nu+1}(z)={\nu \over z}J_{\nu}(z) - {\partial J_{\nu}(z) \over \partial z}
\end{equation}
and the asymptotic limit \cite{STEGUN}

\begin{equation}
\lim_{\nu \rightarrow \infty}J_{\nu}(\nu) \sim 
{2^{1/3} \over 3^{2/3}\Gamma(2/3)\nu^{1/3}},
\end{equation}
giving 

\begin{equation}
\fl R_{1} = {J_{1+2p/r}(2p/r)\over J_{2p/r}(2p/r)} = 1-{\partial \ln J_{\nu}(z)\over \partial z}\bigg|_{z=2p/r}
\!\!\!\!\!\!\!\!\!\!\! \sim 1-{3^{1/3}\Gamma(2/3) \over \Gamma(1/3)}\left({r\over p}\right)^{1/3} \,\mbox{as} \,\,\, p/r \rightarrow \infty.
\end{equation}

\noindent Thus, we find

\begin{equation}
v = p(1-R_{1}) \sim 3^{1/3}{\Gamma(2/3)\over \Gamma(1/3)}r^{1/3}p^{2/3}
\label{VLIMIT}
\end{equation}
in the $r/p\rightarrow 0$ limit. This dimensional result implies that at small
chaperone concentrations, the mean translocation velocity is
proportional to the cube root of the chaperone concentration.
This nonanalytic limit can also be understood physically by considering the 
typical time $t^{*}$ between succesful chaperone depositions approximated by

\begin{equation}
t^{*} \sim {1 \over r x(t^{*})},
\end{equation}
where $x(t^{*})\sim \sqrt{pt^{*}}$ is the typical distance between deposited 
chaperones. Upon solving for $t^{*}$ and estimating the 
velocity as $v \sim x(t^{*})/t^{*} \sim \sqrt{p/t^{*}}$, we find

\begin{equation}
v \sim r^{1/3}p^{2/3}.
\end{equation}

Our results can also be simplified in the limit of infinitely fast
particle deposition, where a particle is deposited as soon as the
first gap reaches a length of $k$ lattice sites. The dynamics then
becomes that of the so called burnt-bridge model
\cite{mai,saffarian,antal,morozov}. In the one-dimensional
burnt-bridge model, certain links (bridges) separated by $k$ lattice
sites can be crossed only once by the walk, which then generates a
biased motion. The speed and the diffusion coefficient can be
calculated by either using the continuous time version of the discrete
time method in \cite{antal}, or by using the general results in
\cite{derrida}, as has been done in \cite{morozov} for an unbiased
($p=q$) random walk. For $r\gg p,q$, we obtain

\begin{equation}
\begin{array}{rl}
\fl  v & \displaystyle =  \frac{pk(a-1)^2}{a(a^k-1)+k(1-a)}\\[13pt]
\fl D & \displaystyle = \frac{pk^2(a-1)^2 \left[ a^{2k+2}+4a^{k+1}(k+1-ka)+k(1-a^2)-a(a+4)\right]}
 {2\left[ a(a^k-1)+k(1-a)\right]^3}
\label{VDR}
\end{array}
\end{equation}

\noindent where $a\equiv q/p$. These expressions simplify further in the 
symmetric case $p=q$:

\begin{equation}
\begin{array}{rl}
 v &  \displaystyle  = \frac{2p}{k+1} \\[13pt]
D & \displaystyle = \frac{2}{3}~\frac{k^2+k+1}{(k+1)^2}p.
\label{VDRSYM}
\end{array}
\end{equation}

\noindent In the large $k$ limit,  appropriate for large binding proteins such as 
single-stranded binding proteins (SSB) with $k \approx 60$, the average velocity and dispersion 
also take on simple limiting forms. For $a=q/p >1$, the large $k$ limit of $v$ and $D$ 
given in Eqs. \ref{VDR} are

\begin{equation}
 v = \frac{pk(a-1)^2}{a^{k+1}} ~,~~~~ D = \frac{pk^2(a-1)^2}{2 a^{k+1}}.
\end{equation}

\noindent Since $q>p$, the drift that tends to close the nearest gap
prevents insertion of particles even if $r$ is large. Thus, both $v$
and $D$ become exponentially small for large $k$.  If $p>q$, the drift
tends to open gaps. However, since very large gaps need to be opened
to allow insertion of a large $k$-site particle, the mean velocity and
dispersion approaches that of a freely diffusing particle: $v = p-q,
\, 2D =p+q$.

\section{Annealing and Zone-Refinement}

Finally, we consider the lattice coverage far from the wall, in the
long time limit. The number of particles adsorbed per length of
translocation may be relevant for considerations about energetics and
macromolecular cost.  Monomers will cover the translocating polymer
behind the wall entirely.  Since deposition has been assumed
irreversible, the deposition of each particle is associated with a
large energy cost, regardless of size.  In this case, monomer
deposition may be more costly than deposition of larger particles, for
which the same energy loss leads to a larger coverage. Particles of
length $k>1$ also allow the presence of empty gaps at saturation,
further minimizing the number of deposited particles.


To quantify a particle cost associated with translocation,
consider the infinite-time coverage $\theta \leq 1$ representing the
fraction of filled lattice sites far behind the moving wall. For
monomers, every site will eventually be filled and $\theta(k=1) =
1$. For $k>1$, we compute $\theta(k)$ by considering the deposition of
a particle into the first gap nearest the wall, splitting it in
two. One of these daughter gaps becomes the new ``first'' gap closest
to the wall, while the other one is now an ``interior'' gap. Particles
will continue to deposit into the interior gaps, following rules of
deposition into a finite length segment \cite{boucher,tchou}, until
these gaps reach a fixed coverage.  For long times we can calculate
the total coverage by summing the saturated coverage of all the
interior gaps. For one particular realization of the random sequential
ratchet, the creation rate of interior gaps of length $m$ from a first
gap of length $m'$ by particle deposition obeys $\dot{N}_{m}(t) = r
H_{m'-m-k}$, where $N_m$ is the number of interior gaps of length
$m$. The Heaviside function prevents the creation of a new interior
gap if the original first gap is not long enough to accommodate the
particle and the new gap.  Ensemble averaging leads to $\langle
\dot{N}_m\rangle = r R_{m+k}$, where $R_{m+k}$ is the probability for
the first gap to be larger than $m+k$ and $\langle N_m\rangle $ is the
ensemble average.  At long times, $R_{m+k}$ reaches steady-state and
the generation of interior gaps grows linearly in time $\langle
N_m(t)\rangle \simeq r t R_{m+k}$.

Interior  gaps created by this process are themselves filled by further
particle deposition. At saturation, an interior gap of initial length $m$
reaches coverage $\theta_{m}$, a quantity that can be calculated by
standard RSA techniques \cite{boucher,tchou}.  Each interior gap is bound
one each end by a particle of length $k$. The total initial length of
one of these interior gaps and an associated end particle is $m+k$. After
further particle deposition into this gap, the number of covered sites
reaches $m \theta_m + k$.  Upon weighting over the ensemble-averaged
gap length distribution $\langle N_{m}\rangle$, the coverage can thus
be expressed as


\begin{equation}
\theta = \frac{\sum_{m=0}^{\infty}(m \theta_m + k) 
R_{m+k}} {\sum_{m=0}^{\infty}(m + k) R_{m+k}}.
\end{equation}


\begin{figure*}
\begin{center}
\includegraphics[height=5.2cm]{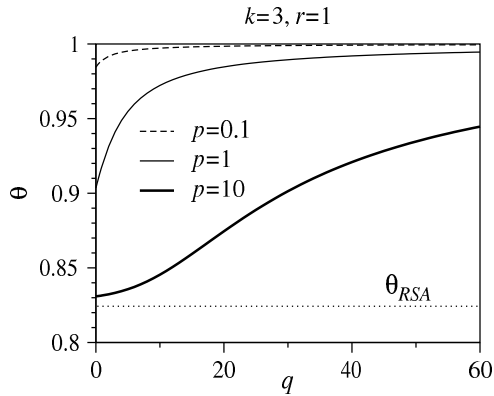}
\end{center}
\caption{Particle coverage $\theta$ behind the wall as a function of $q$ 
for $k=3$, $r=1$, and various $p$.  For walls that move forward rapidly, $p
  \rightarrow \infty$, the coverage corresponds to that of
  irreversible RSA on an infinite lattice without a wall and
  $\theta_{RSA}(k=3)=
  3\sqrt{\pi}e^{-4}(\mbox{Erfi}(2)-\mbox{Erfi}(1))/2= 0.823653$
  (dotted line) For $p \rightarrow 0$, the wall slowly sweeps across
  the lattice, allowing more contiguous deposition behind it. In this
  case $\theta(k, p \rightarrow 0) \sim 1$.}
\label{ZONE}
\end{figure*}

\noindent In Fig.\,\ref{ZONE} we plot $\theta(k=3)$ as a function
of $q$, for $p=0.1,1,10$.  For small $q$ and large $p$, the wall moves
forward rapidly, largely independent of deposition. The depositing
particles rarely interact with the wall and the coverage approaches
that of standard RSA on an infinite lattice, $\theta_{RSA}$
\cite{privman}. If $q$ is large and $p$ is small, the wall stays close
to the nearest particle, occasionally leaving gaps only slightly
larger than $k$ in which particles can deposit.  Thus, the wall slowly
sweeps through the lattice, ``zone-refining'' it by slowing down the
deposition process and allowing for more complete filling. For
intermediate values of $p,q$, we find that the coverage left by the
wall is always between the contiguous and RSA limits.

\section{Summary and Conclusions}

In summary, we have found an exact steady-state velocity of the
discrete translocation problem in the limit of irreversible particle
attachment.  The exact gap distribution $R_{m}$ given by
Eqs.\,\ref{k1sol}, \ref{k2sol}, and \ref{nogo} allows us to construct
the mean velocity from $R_{1}\equiv \langle \sigma\rangle$ and
Eq.\,\ref{ratch}.  Smaller particles yield faster translocation and
smaller dispersion, while larger particles leave less of the remaining
lattice covered.  In the protein translocation problem, neglecting the
dissociation rate as we have done throughout this study is a good
approximation provided chaperone concentrations $>$nM. In this limit,
the result of Elston \cite{ELSTON1} approaches our velocity given in
Eq. \ref{VDRSYM}.  For $p=q$, we find an interesting $r^{1/3}$
dependence of the mean velocity for small $r$. We also find that the
coupling of particle deposition to wall dynamics allows the wall to
``zone-refine'' the lattice, producing long time particle coverages
$\theta_{RSA} \leq \theta \leq 1$.

Our results can be used to guide experimental systems that probe the
mechanisms of chaperone-assisted translocation. For example,
comparison of the mean translocation speed $v$ with our exact solution
and the dispersion $D$ with our Monte-Carlo results, lead to
independent values for $p$ and $q R_{1}$ defined in this paper.  By
measuring $v$ and $D$ for different driving force $F \propto \ln q$
(by tuning {\it e.g.}, a transmembrane potential) one can numerically
determine both the particle size $k$ and the effective adsorption rate
$r$.  If particle detachments occurring at rate $r_{d}$ are also
considered, a sufficient condition for our solution to the mean
velocity $v$ to be accurate is $r_{d} \ll v \theta_{RSA}/k$. For
larger detachment rates, we expect a stall force, where the mean
velocity vanishes, and may be negative for large enough $F$. In this
case, the ratcheting occurs with pawls that detach, allowing
occasional backsliding to the second particle.

\vspace{2mm}

The authors thank Paul Krapivsky, Sidney Redner and Andrej Vilfan for helpful
discussions.  MD and TC were supported by the NSF through grant
DMS-0349195 and by the NIH through grant K25AI058672.  TA acknowledges
financial support to the Program for Evolutionary Dynamics by Jeffrey
Epstein, and NIH grant 1R01GM078986.




\section*{References}
\begin{harvard}

\bibitem[1]{OSTER0} Simon S M,  Peskin CS and  Oster G F 1992 
What drives the translocation of proteins? 
{\it PNAS} {\bf 89} 3770

\bibitem[2]{OSTER1} Peskin C S, Odell G M and  Oster G F 1993
Cellular Motions and Thermal Fluctuations: The Brownian Ratchet
{\it Biophys. J.} {\bf 65} 316


\bibitem[3]{ELSTON1} Elston T C 2002
The Brownain Ratchet and Power Stroke Models for Posttranslational 
Protein Translocation into the Endoplasmic Reticulum 
{\it Biophys. J.} {\bf 82} 1239

\bibitem[4]{gerland}
Gerland U,  Bundschuch R and Hwa T 2004
Translocation of structured polynucleotides through nanopores
{\it Phys. Biol.} {\bf 1} 19

\bibitem[5]{meller}
Meller A,  Nivon L and Branton D 2001
Voltage-Driven DNA Translocations through a Nanopore 
{\it Phys. Rev. Lett.} {\bf 86} 3435

\bibitem[6]{turner}
Turner S W,  Cabodi M and Craighead H G 2002
Confinement-Induced Entropic Recoil of Single DNA Molecules in a Nanofluidic Structure
{\it Phys. Rev. Lett.} {\bf 88} 128103

\bibitem[7]{ZANDI} Zandi R, Reguera D,  Rudnick J and Gelbart W M 2003
What drives the translocation of stiff chains? 
{\it PNAS} {\bf 100} 8649

\bibitem[8]{vale} Vale R D and  Milligan R A 2000
The way things move: looking under the hood
of molecular motor proteins
{\it Science} {\bf 288} 88

\bibitem[9]{RAT0} Simon M and Blobel G 1991 
A protein-conducting channel in the endoplasmic reticulum 
{\it Cell} {\bf 65} 371-380, (1991).

\bibitem[10]{RAT1} Schneider H J,  Berthold J, Bauer M, Dietmeier K,
Guiard B, Brunner M and Neupert W 1994 
Mitochondrial Hsp70/MIM44 complex facilitates protein import 
{\it Nature} {\bf 371} 768

\bibitem[11]{GLICK} Glick B 1995
Can Hsp-70 proteins act as force-generating motors?  
{\it Cell} {\bf 80} 11





\bibitem[12]{KOLO0} Kolomeisky A and Phillips H 2005
Dynamic Properties of Motor Proteins with Two Subunits,
{\it J. Phys. Cond. Matter} {\bf 17} S3887

\bibitem[13]{METZLER} Ambj\"{o}rnsson T, Lomholt M. A. and Metzler R 2005
Directed motion emerging from two coupled random processes:
translocation of a chain through a membrane nanopore driven by binding proteins 
{\it J. Phys.: Condens. Matt.} {\bf 17} 3945

\bibitem[14]{evans} Evans J W 1993
Random and cooperative sequential adsorption 
{\it Rev. Mod. Phys.} {\bf 65} 1281
  
\bibitem[15]{privman} Bartelt M C and Privman V 1991
Kinetics of irrversible monolayer and multilayer adsorption
{\it Int. J. Mod. Phys.} B {\bf 5}
2883

\bibitem[16]{STEGUN} Abramowitz M and Stegun I A, {\it Handbook of Mathematical Functions 
with Formulas, Graphs, and Mathematical Tables}, (Dover Publications, New York, 1964).

\bibitem[17]{jeffreys} H. Jeffreys, {\it Methods of Mathematical
Physics} 3$^{rd}$ed, (Cambridge University Press, 972)

\bibitem[18]{mai} Mai J, Sokolov I M and Blumen A 2001 Directed
particle diffusion under "burnt bridges" conditions {\it Phys. Rev.}
E {\bf 64} 011102

\bibitem[19]{saffarian} Saffarian S, Qian H, Collier I, Elson E and
Goldberg G 2006 Powering a burnt bridges Brownian ratchet: A model for
an extracellular motor driven by proteolysis of collagen {\it
Phys. Rev.} E {\bf 73} 041909


\bibitem[12]{antal} Antal T and Krapivsky P L 2005 ``Burnt-bridge"
mechanism of molecular motor motion {\it Phys.  Rev.} E {\bf 72}
046104

\bibitem[21]{morozov} Morozov A Yu, Pronina E, Kolomeisky A B and
   Artyomov M N 2006 Solutions of burnt-bridge models for molecular
   motor transport Phys. Rev. E {\bf 75} 031910

\bibitem[22]{derrida} Derrida B 1983 Velocity and diffusion constant
of a periodic one-dimensional hopping model {\it J. Stat. Phys.}  {\bf
31} 433

\bibitem[23]{boucher} Boucher E A 1972 
Reaction kinetics of polymer substituents. 
Neighbouring-substituent effects in pairing reactions
{\it J. Chem. Soc. Faraday Trans. 2} {\bf 72} 2281

\bibitem[24]{tchou}  D'Orsogna M R and  Chou T 2005
Gaps on irreversible and equilibrated one-dimensional lattices 
{\it J Phys. A} {\bf 38}, 531-42

\bibitem[25]{watson} Watson G N {\it Treatise of the theory of Bessel functions}. 
2$^{nd}$ed, (Cambridge University Press, 1952).




\end{harvard}
\end{document}